\documentclass[RNAAS]{aastex63}

\usepackage{amsmath}

\begin{document}

\title{A new orbital ephemeris for WASP-128\,b}

\correspondingauthor{Alexis M. S. Smith}
\email{alexis.smith@dlr.de}

\author[0000-0002-2386-4341]{Alexis M. S. Smith}
\affiliation{Institute of Planetary Research, German Aerospace Center (DLR), Rutherfordstra\ss e 2, 12489 Berlin, Germany}

\author[0000-0003-4096-0594]{Philipp Eigm\"uller}
\affiliation{Institute of Planetary Research, German Aerospace Center (DLR), Rutherfordstra\ss e 2, 12489 Berlin, Germany}

\author[0000-0002-0715-8789]{Mahmoudreza Oshagh}
\affiliation{Institut f\"ur Astrophysik, Georg-August Universit\"at G\"ottingen, Friedrich-Hund-Platz 1, 37077, G\"ottingen, Germany}

\author{John Southworth}
\affiliation{Astrophysics Group, Keele University, Staffordshire ST5 5BG, UK}


\section{} 

WASP-128 is a relatively bright ($V = 12.37$) G0-dwarf, known to host a transiting brown dwarf in a short-period orbit \citep{w128}). Very few such objects are known (e.g. \citealt{szilard_corot_book}), which makes WASP-128 a prime target for further observations to better understand giant planet / brown dwarf properties, including formation and migration histories.\\

The only published paper reporting observations of the WASP-128 system to date is H18. The orbital ephemeris reported by H18 is as follows: the time of mid-transit, $T_0 = 2456720.68369 \pm 0.00019$ $\mathrm{BJD_{UTC}}$, and the orbital period, $P = 2.208524^{+0.000021}_{-0.000020}$~d. This gives a 1\,$\sigma$ timing uncertainty of 32 minutes by mid-2020.\\

To facilitate the planning of future observations of WASP-128\,b, we decided to improve the orbital ephemeris of the system, by using the TESS observation of WASP-128. TESS (Transiting Exoplanet Survey Satellite; \cite{TESS}) is currently conducting a photometric survey of around 85 per cent of the sky. WASP-128 was observed in TESS Sector 10 (2019 March 26 to 2019 April 22), but was not a short-cadence target. We therefore downloaded the relevant part of each full-frame image (FFI) from camera 2, CCD 4. We then performed standard aperture photometry on these images.\\

\begin{figure}[h!]
\begin{center}
\includegraphics[scale=0.6,angle=0]{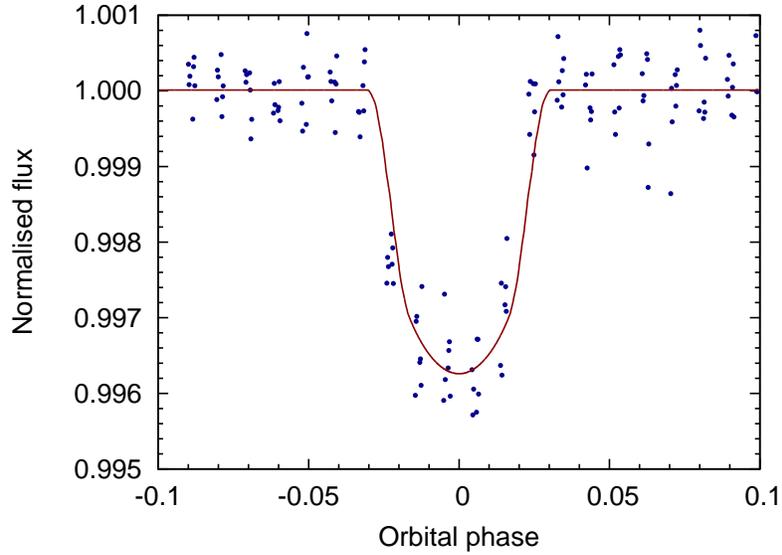}
\caption{Phase-folded TESS light curve of WASP-128, overplotted with our best-fitting model. Note that due to the presence within our photometric aperture of a contaminating third light (with a flux equal to $1.35\pm0.16$ times that of the host star), the transit depth appears suppressed.
\label{fig:1}}
\end{center}
\end{figure}

Seven transits of WASP-128\,b are clearly visible in the resulting light curve. We immediately noticed that the ephemeris of H18 gives a very poor fit to the TESS photometry. Using the TLCM fitting code (\citealt{Szilard_BD}; Csizmadia, in press) we re-fitted the TRAPPIST and EulerCam transit light curves of WASP-128 published by H18, along with the seven transits observed by TESS. The phase-folded TESS data are shown in Fig.~\ref{fig:1}, along with our best-fitting model. Our resulting orbital ephemeris is: 

\begin{gather*}
P = 2.20883665\pm0.00000045~\mathrm{d}\\
T_0 = 2456720.68318 \pm 0.00021~\mathrm{BJD_{UTC}}
\end{gather*}

We note that our orbital period differs significantly from that of H18: it is more than 14\,$\sigma$ (using their uncertainty) larger. This results in our ephemeris predicting a mid-2020 transit to occur almost eight hours later than the H18 ephemeris. This means that anyone using the H18 ephemeris to plan future transit observations would schedule their observations for the wrong night. We find, however, no evidence for any period variation. We tried fitting (i) only the TRAPPIST and EulerCam data, which span 2014 February to 2015 May, and (ii) only the TESS data. In both cases, we found orbital periods compatible with our global fit, and incompatible with that of H18. This strongly suggests that the orbital period published by H18 is erroneous, perhaps as a result of a typographical error. The fact that the period we find is very close to that listed in TEPCat\footnote{\url{https://www.astro.keele.ac.uk/jkt/tepcat/}} \citep{SWorth_homo4}, based on the original WASP data, supports this conclusion. Our newly-determined ephemeris from a fit to TRAPPIST, Euler, and TESS data results in a 1\,$\sigma$ timing error that is just under one minute by mid-2020, and will remain less than ten minutes for the duration of the current century.



\bibliography{refs2}
\bibliographystyle{aasjournal}

\end{document}